\documentclass[runningheads]{llncs}
\usepackage[T1]{fontenc}
\usepackage{graphicx}
\usepackage{amsmath,amssymb,amsfonts}
\usepackage{algorithm}
\usepackage{algorithmic}
\usepackage{xurl} 
\begin{document}
\title{Mal-AAEGAN: Adversarial Autoencoder with Double-Critic WGAN-GP for Malware Generation}
\titlerunning{Mal-AAEGAN for Malware Generation}
\author{Nam Hoang Thanh \and
Trung Pham Duy \and
Lam Bui Thu\thanks{Corresponding author.}}
\authorrunning{N. Hoang Thanh et al.}
\institute{Academy of Cryptography Techniques, Ha Noi, Viet Nam \\
\email{\{hoangthanhnam, trungpd, lambt\}@actvn.edu.vn}}
\maketitle              
\begin{abstract}
Machine-learning (ML) malware detectors are known to be vulnerable to adversarial samples. Generative adversarial networks (GANs) such as MalGAN, Mal-LSGAN, and our previous Mal-D2GAN can craft adversarial malware to probe and harden these detectors, yet they operate directly on a high-dimensional, discrete binary feature space, which leads to unstable training, vanishing gradients, and adversarial samples of limited diversity. To address these problems we propose Mal-AAEGAN. First, an adversarial autoencoder (AAE) learns a low-dimensional, regularized and continuous latent representation of the $M$-dimensional binary API feature vector, and the generator performs an additive perturbation in this latent space rather than in the raw binary space. Second, training follows the Wasserstein GAN with gradient penalty (WGAN-GP) scheme using two critics: a mimic-black-box critic that inherits the role of the substitute detector and a naive critic that inherits the role of the additional detector. A decode-binarize-OR pipeline guarantees that the original API calls of the malware are preserved. The model is evaluated on two malware datasets of different input dimensionality ($M\in\{139,215\}$) against eight common ML detectors and compared, under an identical protocol, with MalGAN, Mal-LSGAN and Mal-D2GAN. Mal-AAEGAN trains stably in the continuous latent space and attains an evasion rate on par with or higher than these binary-space baselines across both datasets, while its generator stays responsive to the noise input rather than reducing to a single fixed perturbation per malware.

\end{abstract}
\section{Introduction}
Malware refers to programs intentionally designed to harm computers, servers, or networks. Detecting malware is a critical step in the incident-response process of any organization. Detection approaches fall into two families: signature-based detection, which compares files against a database of known byte patterns, and behavior-based detection, which analyses program actions using heuristic rules or machine-learning algorithms and can flag previously unseen samples \cite{b1,b2}.

Machine learning learns from program behavior instead of relying only on known signatures \cite{b3}, enabling the detection of new malware variants. However, ML-based detectors are themselves susceptible to adversarial samples \cite{b4}. Samples produced by generative adversarial networks (GANs) \cite{b5} have driven numerous attacks against such detectors, as surveyed in \cite{b32}. Hu et al. introduced MalGAN, which trains a generator against a substitute detector so that the resulting samples bypass a black-box detector \cite{b6}. Mal-LSGAN and LSGAN-AT later improved evasion through least-square losses and new activation functions \cite{b7,b8}, while our earlier Mal-D2GAN \cite{b31} added a second detector to strengthen the adversarial signal.

These models share a structural limitation: the generator operates on the raw $M$-dimensional binary feature vector. Because this space is discrete and high-dimensional, the gradient passed back to the generator is weak, training is unstable, and samples lack diversity. In this paper we revisit the Mal-D2GAN design and rebuild its generator and loss around a continuous latent space. The contributions of this work are:
\begin{enumerate}
	\item \textit{The Mal-AAEGAN model.} We propose Mal-AAEGAN, which couples an adversarial autoencoder that maps the $M$-dimensional binary API features into a compact, continuous and prior-regularized latent space with a double-critic WGAN-GP trained inside that space. A single generator applies an additive perturbation to the malware latent code and is optimized by the negative average of two Wasserstein critic scores, a mimic-black-box critic and a naive critic, each regularized by a gradient penalty. This objective provides smooth gradients and stable training while retaining the double-detector signal of Mal-D2GAN.
	\item \textit{Noise-responsive latent perturbations.} Combining the regularized latent space with the WGAN-GP objective provides smooth gradients and stable training in the malware domain. We introduce dedicated diversity metrics, unique ratio, mean Hamming distance, latent-dispersion ratio, and noise sensitivity, and evaluate the model on two datasets against eight detectors. Mal-AAEGAN attains evasion on par with the binary-space baselines at a comparable perturbation cost, and its generator remains responsive to its noise input where the baselines effectively ignore it.
\end{enumerate}
The remainder of the paper is organized as follows. Section~\ref{sec:related} reviews related work. Section~\ref{sec:method} presents the proposed Mal-AAEGAN. Section~\ref{sec:exp} reports the experiments and results, and Section~\ref{sec:conc} concludes the paper.

\section{Related Work}\label{sec:related}
\textit{MalGAN.} Hu et al. proposed MalGAN, the first GAN-based generator of adversarial malware able to evade black-box ML models \cite{b6}. A substitute detector is fitted to the black-box decisions and the generator is trained to minimize the malware probability predicted by this substitute. MalGAN can drive the detection rate to nearly zero, which also weakens retraining-based defenses.

\textit{Mal-LSGAN and LSGAN-AT.} To stabilize training, Mal-LSGAN replaces the binary cross-entropy loss with a least-square loss and combines new activation functions, achieving a higher attack success rate and a lower true-positive rate (TPR) across six detectors \cite{b7,b8,b9}.

\textit{AAGAN.} AAGAN is an automated Android-malware generation system that flips the prediction of state-of-the-art detectors in up to 99\% of cases; after five retraining rounds it still reaches an 89\% evasion rate \cite{b10}.

\textit{Adversarial attacks on mobile malware.} Shahpasand et al. bound the number of changes to an original sample and, on the Drebin dataset, evade four classifier families with up to a 99\% success rate \cite{b12}.

\textit{Mal-D2GAN.} Our previous model \cite{b31} added an additional detector that works together with the substitute detector, reducing the TPR of eight detectors close to zero. Mal-AAEGAN keeps this double-detector philosophy but rebuilds the generator and the loss around an AAE latent space and the WGAN-GP objective.

\textit{Generative models for malware augmentation.}
A parallel line of work applies generative models defensively, to enrich scarce datasets rather than to evade detectors. Bao et al.~\cite{b30} decompose binaries into mnemonic opcode sequences, embed each opcode with Word2Vec and average them into a single per-file vector, then compare a GAN, a WGAN-GP and a 1-D U-Net diffusion model. Augmenting the training set with the generated embeddings raises minority-family accuracy by up to $60\%$ and the overall accuracy to $96\%$. Their results support two premises of our design, namely that a continuous space is far more tractable for generative training than the raw discrete one, and that WGAN-GP is the reference objective in this domain.

The threat models nonetheless differ: their embeddings are sampled unconditionally, are never inverted into a valid feature vector and never face a black-box detector, so neither functionality preservation nor evasion applies. Mal-AAEGAN instead perturbs an existing malware code inside a prior-regularized AAE latent space and returns it to a valid $M$-dimensional binary vector through the decode - binarize - OR pipeline, under critics driven by black-box labels.

\textit{Latent-space and autoencoder-based generation.} A recent line of work moves adversarial generation off the raw binary space into a learned latent representation. Ajayi \textit{et~al.}~\cite{b34} build variational-autoencoder latent spaces for malware modelling and discuss hybrid adversarial--variational autoencoders that combine an AAE with a VAE to capture discriminative structure in the latent code. Closest to our setting, Dautartas \textit{et~al.}~\cite{b33} employ a conditional VAE whose decoder is \emph{strictly additive}: it injects benign-looking Win32 API imports without removing any existing call, preserving malware functionality by design, and reports transfer to commercial engines on VirusTotal. Mal-AAEGAN shares this latent-space and additive, functionality-preserving philosophy, but replaces the (C)VAE generator with an AAE-regularized manifold trained under a \emph{double-critic} WGAN-GP objective driven by black-box labels, which additionally targets the noise-response limitation that single-generator models do not address.

\textit{Foundations.} The adversarial autoencoder \cite{b22} regularizes the latent posterior to match an arbitrary prior with an adversarial discriminator, yielding a smooth latent manifold. WGAN \cite{b23} and WGAN-GP \cite{b24} minimize the Wasserstein distance and enforce a 1-Lipschitz critic through a gradient penalty, which improves training stability over the original GAN. We combine both ideas in the malware domain.

\section{The Proposed Mal-AAEGAN}\label{sec:method}
\begin{figure}[t]
	\centering
	\includegraphics[width=\textwidth]{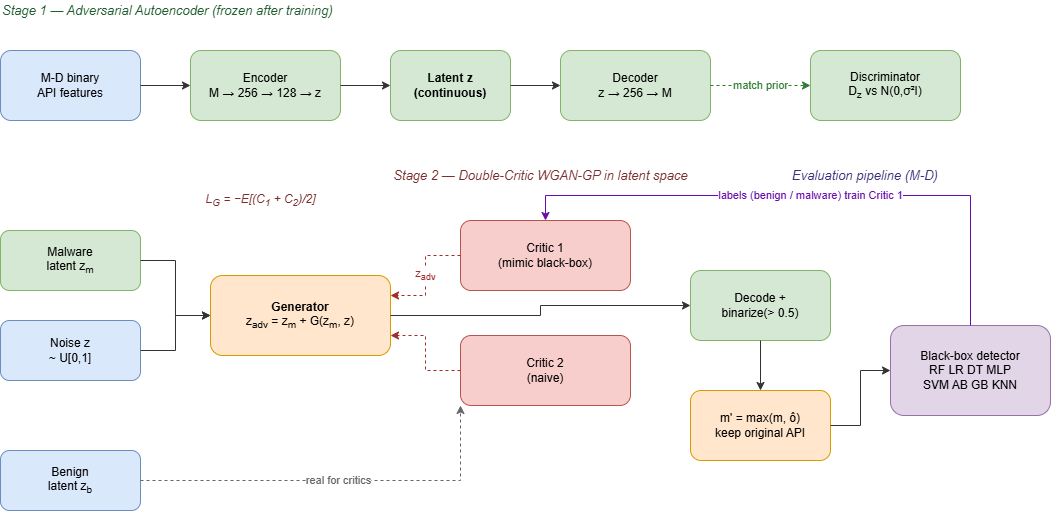}
	\caption{Architecture of Mal-AAEGAN. Stage 1 (top): the AAE encodes the $M$-D binary features into a regularized latent space. Stage 2 (bottom): the generator perturbs malware latent codes and is trained with WGAN-GP against two critics. The black-box detector always evaluates $M$-D binary samples produced by the decode-binarize-OR pipeline.}
	\label{fig:arch}
\end{figure}

Mal-AAEGAN is built in two stages. In the first stage an adversarial autoencoder learns a continuous latent representation of the malware feature space and is then frozen. In the second stage a double-critic WGAN-GP is trained inside that latent space. Throughout, the black-box detector always operates on the original $M$-dimensional binary vectors, where $M$ is the number of API features of the dataset ($M=139$ and $215$ for CICMaldroid 2020  and Drebin, respectively); the latent space is used only by the generator and the critics. Fig.~\ref{fig:arch} shows the overall architecture.

\subsection{Black-box detector}
As in Mal-D2GAN, the black-box detector is an external ML system. The attacker knows only the general features it consumes, not its algorithm or parameters, but can query its decisions. Eight classifiers are used: Random Forest (RF), Logistic Regression (LR), Decision Tree (DT), Multilayer Perceptron (MLP), Support Vector Machine (SVM), AdaBoost (AB), Gradient Boosting (GB) and K-Nearest-Neighbor (KNN).

\subsection{Adversarial autoencoder (latent representation)}
The AAE \cite{b22} consists of three sub-networks. The \textit{encoder} maps an $M$-dimensional input $x$ to a latent code $z$ through 256 and 128 hidden units with ReLU activations and a linear output. The \textit{decoder} reconstructs $x$ from $z$ through 128 and 256 units back to $M$ dimensions with a sigmoid output. A \textit{latent discriminator} $D_z$ tries to separate codes drawn from a Gaussian prior $\mathcal{N}(0,\sigma^2 I)$ from codes produced by the encoder.

Each minibatch is trained in three phases. The reconstruction phase minimizes the binary cross-entropy between $x$ and its reconstruction:
\begin{equation}
	L_{\mathrm{recon}} = \mathrm{BCE}\big(x,\; \mathrm{Dec}(\mathrm{Enc}(x))\big). \label{eq:recon}
\end{equation}
The adversarial phases then align the aggregated posterior $q(z)$ with the prior. $D_z$ is updated to classify prior samples as real and encoder samples as fake, and the encoder is updated to fool $D_z$:
\begin{align}
	L_{D_z} &= -\mathbb{E}\big[\log D_z(z_{\mathrm{prior}})\big] \nonumber\\
	&\quad -\mathbb{E}\big[\log\big(1 - D_z(\mathrm{Enc}(x))\big)\big], \label{eq:dz}\\
	L_{\mathrm{enc}} &= -\mathbb{E}\big[\log D_z(\mathrm{Enc}(x))\big]. \label{eq:enc}
\end{align}
This adversarial regularization yields a smooth and continuous latent manifold, which is the key enabler for stable gradient flow in the second stage. After training, the encoder and decoder are frozen and reused by the generator and the evaluation pipeline.

\subsection{Latent-dimension selection}\label{sec:latdim}
The latent dimension controls the trade-off between compactness and fidelity. We first run a PCA baseline on the combined data to estimate a reasonable range, then perform an ablation over the candidate dimensions $\{8,16,24,32,48,64,96\}$. For each candidate we train a short AAE and measure four metrics: (M1) bit reconstruction accuracy; (M2) the F1 score on active bits, which accounts for the sparsity of the feature vectors; (M3) semantic preservation, defined as the accuracy ratio of a Random-Forest classifier on reconstructed versus original test data; and (M4) the accuracy of $D_z$, which is best when close to 0.5 because that indicates $q(z)$ matches the prior. Candidates must pass three gates ($\mathrm{M1}\geq 0.95$, $\mathrm{M3\ ratio}\geq 0.90$, $\mathrm{M4}\leq 0.65$); among the qualified candidates we adopt the most compact dimension that also attains the best reconstruction--regularization balance. On both datasets the most compact candidate $d{=}16$ already passes all three gates, so we adopt $d=16$: on CICMaldroid 2020 it reaches M1$=0.995$, M3$=0.966$ and a $D_z$ accuracy of $0.599$, and on Drebin M1$=0.984$, M3$=0.991$ and a $D_z$ accuracy of $0.5104$. Because a more compact code that clears every gate is preferable, no larger dimension is needed. Table~\ref{tab:ablation} reports the four fidelity metrics at the selected dimension.

\subsection{Generator}
The generator takes a malware latent code $z_m$ and a noise vector $z\sim U[0,1)$ and outputs an \textit{additive} perturbation in the latent space:
\begin{equation}
	z_{\mathrm{adv}} = z_m + G(z_m, z). \label{eq:gen}
\end{equation}
Because the perturbation is added to $z_m$, the adversarial code starts close to the original malware and the network only has to learn the residual that fools the critics. The generator uses two hidden layers (256 and 128 units) with LeakyReLU activations and a linear output of the latent dimension.

\subsection{Double critic and WGAN-GP objective}
Two critics replace the substitute and additional detectors of Mal-D2GAN. Each critic is a latent-space network with LeakyReLU layers, no batch normalization, and a single linear output (no sigmoid), as required by WGAN-GP.

\textit{Critic 1 (mimic black-box)} inherits the role of the substitute detector. Its real samples are benign latent codes that the black-box labels as benign, and its fake samples are generated latent codes that the black-box labels as malware. Its loss is
\begin{equation}
	L_{C_1} = \mathbb{E}\big[C_1(z_{\mathrm{fake}})\big] - \mathbb{E}\big[C_1(z_{\mathrm{real}})\big] + \lambda\,\mathrm{GP}. \label{eq:c1}
\end{equation}
\textit{Critic 2 (naive)} inherits the role of the additional detector. It treats real benign latent codes as real and every generated code as fake, without querying the black-box:
\begin{equation}
	L_{C_2} = \mathbb{E}\big[C_2(z_{\mathrm{fake}})\big] - \mathbb{E}\big[C_2(z_{\mathrm{real}})\big] + \lambda\,\mathrm{GP}. \label{eq:c2}
\end{equation}
In both equations GP is the gradient penalty that enforces a 1-Lipschitz critic on points $\hat{z}$ interpolated between real and fake codes:
\begin{equation}
	\mathrm{GP} = \mathbb{E}\Big[\big(\lVert \nabla_{\hat{z}} C(\hat{z}) \rVert_2 - 1\big)^2\Big],\;\; \hat{z}=\epsilon\, z_{\mathrm{real}} + (1-\epsilon)z_{\mathrm{fake}}. \label{eq:gp}
\end{equation}
with $\epsilon\sim U[0,1]$ drawn independently per sample. A single generator must fool both critics. Mirroring the weighted double-detector loss of Mal-D2GAN, the generator minimizes the negative weighted average critic score:
\begin{equation}
	L_{G} = -\mathbb{E}\big[\alpha\,C_1(z_{\mathrm{adv}}) + (1-\alpha)\,C_2(z_{\mathrm{adv}})\big], \label{eq:g}
\end{equation}
where $\alpha\in[0,1]$ is the critic importance weight; $\alpha=0.5$ recovers the plain average shown in Fig.~\ref{fig:arch}. Following WGAN-GP practice, each critic is updated $n_{\mathrm{critic}}$ times per generator update, the penalty coefficient is $\lambda=10$, and Adam uses $\beta_1=0.5$, $\beta_2=0.9$.

\subsection{Adversarial-sample pipeline}
Critics operate in the latent space, but the black-box detector must receive a valid $M$-D binary vector. Every adversarial code is therefore decoded, binarized at threshold 0.5, and combined with the original malware vector through an element-wise OR so that no original API call is removed:
\begin{equation}
	m' = \max\big(m,\; \mathrm{binarize}(\mathrm{Dec}(z_{\mathrm{adv}}))\big). \label{eq:pipe}
\end{equation}
Keeping the original bits guarantees that the malware retains its functionality; the generator can only add unrelated features, exactly as in the binary-space formulation of Mal-D2GAN. The resulting $m'$ is the sample submitted to the black-box detector during both training (for the Critic-1 labels) and evaluation.

\subsection{Training procedure}
Algorithm~\ref{alg:train} summarizes one training epoch of the second stage. The AAE is assumed pre-trained and frozen.

\begin{algorithm}[t]
	\caption{Training of Mal-AAEGAN (second stage)}
	\label{alg:train}
	\begin{algorithmic}[1]
		\REQUIRE frozen AAE $(\mathrm{Enc},\mathrm{Dec})$, black-box $B$
		\WHILE{not converged}
		\FOR{$n_{\mathrm{critic}}$ steps}
		\STATE sample malware $z_m$, benign $z_b$, noise $z$
		\STATE $z_{\mathrm{adv}} \leftarrow z_m + G(z_m,z)$
		\STATE $m' \leftarrow \max(m,\mathrm{binarize}(\mathrm{Dec}(z_{\mathrm{adv}})))$
		\STATE label $m'$ and benign with $B$; select $C_1$ real/fake
		\STATE update $C_1$ by $\nabla L_{C_1}$; update $C_2$ by $\nabla L_{C_2}$
		\ENDFOR
		\STATE update $G$ by $\nabla L_{G}$ (fool $C_1$ and $C_2$)
		\ENDWHILE
	\end{algorithmic}
\end{algorithm}

\section{Experiments and Results}\label{sec:exp}
\subsection{Experimental setup}
We evaluate Mal-AAEGAN on two malware feature datasets that differ in origin and input dimensionality $M$:

(i)~\textit{CICMaldroid 2020} ($M=139$), whose samples are obtained by selecting dynamic features, giving approximately $9{,}803$ malware and $1{,}795$ benign samples described by $139$ features.

(ii)~\textit{Drebin} ($M=215$), a widely used Android malware feature set~\cite{b28} whose samples are described by $215$ binary static features (permissions, API calls, intents, and hardware components).

Each dataset is split $80/20$ into training and test sets, and the critic importance weight in \eqref{eq:g} is set to $\alpha=0.5$. Unless otherwise stated, the tables in this section report the CICMaldroid 2020 run; Drebin is evaluated under the identical two-stage protocol.
Experiments ran on an Intel Core i5-12500H with 16 GB RAM under Windows 11 and Anaconda \cite{b17}. The AAE is trained for 200 epochs; the WGAN-GP stage uses 200 epochs with $n_{\mathrm{critic}}=5$, $\lambda=10$, learning rates $10^{-4}$, and batch size 64.

We report the true-positive rate (TPR) \cite{b19}, i.e. the fraction of malware correctly detected, and the evasion rate $\mathrm{ER}=1-\mathrm{TPR}$. A lower TPR (higher ER) on adversarial samples means a stronger attack.

\subsection{Evaluation metrics}\label{sec:metrics}
We evaluate the generated samples along three axes: (i)~\textit{detection evasion}, following the black-box protocol of MalGAN~\cite{b6}; (ii)~\textit{perturbation cost and functionality validity}, following the additive-only constraint of Grosse~\textit{et~al.}~\cite{b25}; and (iii)~\textit{sample diversity}, for which we adapt established generative-model diagnostics~\cite{b27,b26} to the binary feature space. To keep the contribution transparent, we label each metric as \emph{standard} (adopted from prior work), \emph{adapted} (re-expressed for the binary malware space), or \emph{ours} (a diagnostic operationalized in this work). Let $m\in\{0,1\}^{M}$ (with $M$ the input dimensionality of the dataset, $M\in\{215,139\}$) be an original malware vector, $m'$ its adversarial counterpart from \eqref{eq:pipe}, $B(\cdot)$ the black-box detector ($1$=malware, $0$=benign), $N$ the number of malware samples, and $d$ the latent dimension.

\textit{Latent-space fidelity.} The four ablation metrics (M1--M4) used in Table~\ref{tab:ablation} are formalized as follows. With the binarized reconstruction $\hat{x}=\mathbb{1}[\mathrm{Dec}(\mathrm{Enc}(x))>0.5]$, the bit accuracy and the semantic-preservation ratio are
\begin{equation}
	\mathrm{M1}=\frac{1}{NM}\sum_{i,j}\mathbb{1}[\hat{x}_{ij}=x_{ij}],\quad
	\mathrm{M3}=\frac{\mathrm{Acc}_{\mathrm{RF}}(\hat{X}_{\mathrm{test}})}{\mathrm{Acc}_{\mathrm{RF}}(X_{\mathrm{test}})}, \label{eq:m13}
\end{equation}
\begin{equation}
	\mathrm{M2}=\mathrm{F1}\big|_{x_{ij}=1},\qquad
	\mathrm{M4}=\tfrac{1}{2}\big(\mathrm{Acc}_{D_z}(z_{\mathrm{prior}})+\mathrm{Acc}_{D_z}(q(z))\big), \label{eq:m24}
\end{equation}
where M2 is the F1 score of the reconstruction restricted to the active bits ($x_{ij}{=}1$), which accounts for feature sparsity, and M4 (best near $0.5$) is the mean accuracy of $D_z$ over prior samples $z_{\mathrm{prior}}$ and posterior codes $q(z)=\mathrm{Enc}(x)$.

\textit{Attack effectiveness (standard~\cite{b6}).} The TPR on adversarial samples and the evasion rate are
\begin{equation}
	\mathrm{TPR}=\frac{1}{N}\sum_{i=1}^{N}\mathbb{1}[B(m'_i)=1],\quad \mathrm{ER}=1-\mathrm{TPR}. \label{eq:er}
\end{equation}
A stronger attack yields a lower TPR (higher ER)~\cite{b6}.

\textit{Perturbation cost and validity (standard~\cite{b25}).} Under the additive-only constraint that preserves malicious functionality~\cite{b25}, the number of API features added to sample $i$ (an $\ell_1$ distance in the binary space) and its mean $\bar a$, the average number of added API features that we report as the perturbation cost, are
\begin{equation}
	a_i=\sum_{j=1}^{M}(m'_{ij}-m_{ij}),\quad \bar a=\frac{1}{N}\sum_{i=1}^{N}a_i. \label{eq:cost}
\end{equation}
Functionality preservation is checked by the total removed-bit count $R$ and the valid-sample ratio $V$:
\begin{equation}
	R=\sum_{i=1}^{N}\sum_{j=1}^{M}\mathbb{1}[m_{ij}{=}1\wedge m'_{ij}{=}0],\quad V=\frac{1}{N}\sum_{i=1}^{N}\mathbb{1}[r_i{=}0], \label{eq:valid}
\end{equation}
where $r_i=\sum_j\mathbb{1}[m_{ij}{=}1\wedge m'_{ij}{=}0]$. Because $m'=\max\!\big(m,\mathrm{binarize}(\mathrm{Dec}(z_{\mathrm{adv}}))\big)$, the OR pipeline guarantees $R=0$ and $V=1$; both are reported as a sanity check that no original API call is removed.

\textit{Diversity.} The unique ratio $U$ (\emph{adapted}: a discrete-space analogue of the birthday-paradox support-size test~\cite{b26}) and the mean pairwise normalized Hamming distance $H$ (\emph{ours}: an intra-set diversity measure for binary features) quantify inter-sample diversity:
\begin{equation}
	U=\frac{|\{m'_i\}_{i=1}^{N}|}{N},\quad H=\frac{2}{N(N{-}1)}\sum_{i<j}\frac{1}{M}\sum_{k=1}^{M}|m'_{ik}-m'_{jk}|. \label{eq:div}
\end{equation}
In the latent space, the dispersion ratio $\rho$ and the noise sensitivity $S_{\mathrm{noise}}$ (both \emph{ours}, following the noise-response diagnostic of~\cite{b27}) are
\begin{align}
	\rho &= \Big(\tfrac{1}{d}\textstyle\sum_{l=1}^{d}\operatorname{std}_i(z_{\mathrm{adv},il})\Big) \Big/ \Big(\tfrac{1}{d}\textstyle\sum_{l=1}^{d}\operatorname{std}_i(z_{m,il})\Big), \label{eq:rho}\\
	S_{\mathrm{noise}} &= \mathbb{E}_{z_m}\Big[\tfrac{1}{d}\textstyle\sum_{l=1}^{d}\operatorname{std}_{t}\big(G(z_m,z^{(t)})_l\big)\Big], \label{eq:snoise}
\end{align}
where the inner standard deviation is computed for a fixed $z_m$ over $T$ noise draws $z^{(t)}$ and then averaged over the malware set. A higher unique ratio indicates more distinct adversarial samples, $\rho$ compares the spread of the generated codes with that of the original malware codes, and $S_{\mathrm{noise}}\approx 0$ means the generator effectively ignores its noise input.

\begin{table}[htbp]
	\caption{Latent-space fidelity of the AAE at the selected latent dimension per dataset.}
	\label{tab:ablation}
	\centering
	\begin{tabular}{|c|c|c|c|c|c|}
		\hline
		\textbf{Dataset ($M$)} & \textbf{$d$} & \textbf{M1 bit acc.} & \textbf{M2 F1} & \textbf{M3 ratio} & \textbf{M4 $D_z$} \\
		\hline
		CICMaldroid 2020 (139)& 16 & 0.995 & 0.995 & 0.966 & 0.599 \\\hline
		Drebin (215)   & 16 & 0.984 & 0.948 & 0.991 & 0.510 \\\hline
	\end{tabular}
	\vspace{4pt}

	{\footnotesize Note: at the per-dataset selected $d$ ($16$ for both CICMaldroid 2020 and Drebin). M1 = bit reconstruction accuracy (gate $\geq0.95$); M2 = F1 on active bits; M3 = RF accuracy ratio reconstructed/original (gate $\geq0.90$; $0.966$ and $0.991$ for CICMaldroid 2020/Drebin); M4 = $D_z$ accuracy, best near $0.50$ (gate $\leq0.65$). Both datasets pass every gate.}
\end{table}

Table~\ref{tab:ablation} reports the four latent-space fidelity metrics at the per-dataset selected dimension ($d=16$ for both CICMaldroid 2020 and Drebin); all of them clear the three acceptance gates on both datasets, which justifies freezing one AAE configuration per dataset before the attack stage. Reconstruction quality (M1) exceeds $0.98$ on both datasets ($0.9949$ on CICMaldroid 2020, $0.984$ on Drebin), and the F1 on active bits stays high ($0.9946$ and $0.948$), confirming that the reconstruction criterion is met despite the sparsity of the binary features. Semantic preservation (M3) is high on both datasets ($0.966$ on CICMaldroid 2020 and $0.991$ on Drebin): a Random-Forest classifier loses at most about $3\%$ accuracy when moved from original to reconstructed data, so the encoder--decoder retains the class-discriminative structure on which the downstream attack relies. Finally, the discriminator accuracy (M4) stays within the $0.65$ gate on both datasets ($0.599$ on CICMaldroid 2020, $0.510$ on Drebin): the discriminator is only moderately better than chance at separating the aggregated posterior $q(z)$ from the Gaussian prior, so the regularization holds on both feature spaces. Overall, the selected dimension yields a compact latent space (down from the $139$ and $215$ input dimensions) that is simultaneously reconstructive, semantics-preserving and well-regularized, which is the prerequisite for the stable latent-space WGAN-GP training of the next stage.

\subsection{Detection results}
Tables~\ref{tab:test} and~\ref{tab:testnew} report, per detector, the evasion rate of all four generators on CICMaldroid 2020 and Drebin; every model was trained and evaluated under the identical protocol on the same features. A successful attack drives the evasion rate toward~$100\%$.

\begin{table}[t]
	\caption{Evasion rate (\%) of the four generators per detector, CICMaldroid 2020 ($M{=}139$). Higher is a stronger attack.}
	\label{tab:test}
	\centering
	\begin{tabular}{|c|c|c|c|c|c|}
		\hline
		\textbf{Detector} & \textbf{Orig.\ TPR} & \textbf{Mal-AAEGAN} & \textbf{MalGAN}& \textbf{Mal-D2GAN}& \textbf{Mal-LSGAN}\\
		\hline
		DT& 98.72&  \textbf{98.93}&  93.5&  93.0&  95.32\\\hline
		KNN& 99.56&  94.43 &  98.0&  92.9&  98.83\\\hline
		GB& 98.76& \textbf{96.20}&  90.86&  95.7&  88.74\\\hline
		AB& 97.80&  \textbf{99.83}&  97.23&  96.8&  97.18\\\hline
		MLP& 99.26& 99.85 &  100&  95.4& 99.99\\\hline
		RF& 98.91&\textbf{  99.43}&  97.75&  86.2& 97.91\\\hline
		SVM& 98.69&  99.62& 99.77&  97.9&  99.85\\\hline
		LR& 98.78&  99.84& 100&  92.9&  99.91\\\hline
	\end{tabular}
	\vspace{4pt}

	{\footnotesize Note: all four generators trained and evaluated under the identical protocol on the same $139$ features (round-1 attack against the original black-box). Mal-AAEGAN at $d{=}16$.}
\end{table}

\begin{table}[t]
	\caption{Evasion rate (ER) (\%) of the four generators per detector, Drebin ($M{=}215$). Higher is a stronger attack.}
	\label{tab:testnew}
	\centering
	\begin{tabular}{|c|c|c|c|c|c|}
		\hline
		\textbf{Detector} & \textbf{Orig.\ TPR} & \textbf{Mal-AAEGAN} & MalGAN& Mal-D2GAN& Mal-LSGAN\\
		\hline
		DT& 96.49 & 91.3& 99.9& 96.0& 99.8\\\hline
		KNN& 97.93 & \textbf{100}& 99.9& 97.6& 99.7\\\hline
		GB& 97.30 & \textbf{100}& 95.6& 98.8& 100\\\hline
		AB& 93.62 & \textbf{100}& 100& 99.6& 100\\\hline
		MLP& 96.94 & \textbf{100}& 89.10& 99.6& 100\\\hline
		RF& 96.94 & \textbf{100}& 94.0& 99.8& 99.9\\\hline
		SVM& 96.58 &\textbf{ 100}& 86.6& 99.6& 100\\\hline
		LR & 96.22 & \textbf{100}& 91.7& 99.6& 100\\\hline
	\end{tabular}
	\vspace{4pt}

	{\footnotesize Note: identical protocol on the same $215$ features. Mal-AAEGAN at $d{=}16$.}
\end{table}

On CICMaldroid 2020, Mal-AAEGAN reaches an evasion rate of $94.4$--$99.9\%$ across the eight detectors (mean $98.5\%$; the adversarial TPR drops to $0.2$--$5.6\%$); on Drebin the evasion rate is $91.3$--$100\%$ (mean $98.9\%$), reaching $100\%$ on seven of the eight detectors. On both datasets this evasion is on par with or above the binary-space baselines (Table~\ref{tab:modelcmp}), at a comparable perturbation cost. We examine the per-model comparison next.

\subsection{Sample quality and diversity}
Table~\ref{tab:modelcmp} compares the four generators on perturbation cost \eqref{eq:cost} and diversity \eqref{eq:div}--\eqref{eq:snoise}, averaged over the eight detectors. \emph{Cost:} the four models add a similar number of API features -- between $20.7$ and $51.1$ features on CICMaldroid 2020 and between $42.6$ and $91.9$ on Drebin. Mal-AAEGAN adds $34.2$ and $42.6$ features on average, respectively: the fewest of all models on Drebin, and fewer than MalGAN and Mal-LSGAN on CICMaldroid 2020 (only Mal-D2GAN adds fewer there, $20.7$). Because the OR pipeline \eqref{eq:pipe} never removes an original call, every sample of every model stays valid ($R=0$, $V=1$). \emph{Diversity:} Mal-AAEGAN is the only generator whose noise sensitivity is non-negligible ($0.075$ on CICMaldroid 2020 and $0.411$ on Drebin), whereas the baselines sit at $S_{\mathrm{noise}}\approx0$ and map each malware to essentially one adversarial output. Its unique ratio is moderate on both datasets ($0.320$ on CICMaldroid 2020 and $0.799$ on Drebin). Tables~\ref{tab:collapse} and~\ref{tab:div215} give the per-detector diversity indicators \eqref{eq:div}--\eqref{eq:snoise} for Mal-AAEGAN: the unique ratio ranges from $0.07$ to $0.51$ on CICMaldroid 2020 and from $0.61$ to $1.00$ on Drebin.

\begin{table}[t]
	\caption{Model comparison averaged over the eight detectors: evasion (ER), perturbation cost (average number of added API features, Add) and diversity (unique ratio $U$, noise sensitivity $S_{\mathrm{noise}}$). Best per column in bold. Mal-AAEGAN attains the strongest evasion on CICMaldroid 2020 and a competitive evasion on Drebin at a low perturbation cost, and is the only model with a non-negligible $S_{\mathrm{noise}}$.}
	\label{tab:modelcmp}
	\centering
	\resizebox{\textwidth}{!}{%
	\begin{tabular}{|l|cccc|cccc|}
		\hline
		 & \multicolumn{4}{c|}{\textbf{CICMaldroid 2020 ($M{=}139$)}} & \multicolumn{4}{c|}{\textbf{Drebin ($M{=}215$)}} \\
		\cline{2-9}
		\textbf{Model} & ER (\%) & Avg added API& $U$ & $S_{\mathrm{noise}}$ & ER (\%) & Avg added api& $U$ & $S_{\mathrm{noise}}$ \\
		\hline
		Mal-AAEGAN & \textbf{98.51}& 34.2 & 0.320 & \textbf{0.075} & 98.91& \textbf{42.61}& 0.799& \textbf{0.456}\\\hline
		Mal-D2GAN & 93.84& \textbf{20.71}& 0.25& 0.0008& 98.96& 51.53& 0.6929& 0.0542\\\hline
		Mal-LSGAN & 97.21& 46.6& \textbf{0.57}& 0.011& \textbf{99.93}& 91.94& 0.929 & 0.023 \\\hline
		MalGAN & 97.26& 51.12& 0.1979& 0.0002& 94.6& 62.84& \textbf{0.936}& 0.0094\\\hline
	\end{tabular}%
	}
	\vspace{4pt}

	{\footnotesize Note: ER $=1-\mathrm{TPR}$ (higher $=$ stronger evasion); Add is the average number of API features (calls) added per sample. $R=0$, $V=1$ for every model. $S_{\mathrm{noise}}\approx0$ means the generator ignores its noise input.}
\end{table}

\begin{table}[t]
	\caption{Sample diversity indicators (round-1 attack), CICMaldroid 2020 ($M{=}139$).}
	\label{tab:collapse}
	\centering
	\begin{tabular}{|c|c|c|c|c|}
		\hline
		\textbf{Detector} & \textbf{$U$} & \textbf{$H$} & \textbf{$\rho$} & \textbf{$S_{\mathrm{noise}}$} \\
		\hline
		DT & 0.25 & 0.06 & 1.95 & 0.04\\\hline
		KNN & 0.31 & 0.08 & 2.25 & 0.03\\\hline
		GB & 0.24 & 0.06 & 1.16 & 0.05\\\hline
		AB & 0.51 & 0.08 & 2.14 & 0.06\\\hline
		MLP & 0.45 & 0.08 & 1.52 & 0.08\\\hline
		RF & 0.07 & 0.01 & 1.94 & 0.14\\\hline
		SVM & 0.42 & 0.06 & 0.78 & 0.05\\\hline
		LR & 0.29 & 0.04 & 2.25 & 0.14\\\hline
	\end{tabular}
	\vspace{4pt}

	{\footnotesize Note: $U$=unique ratio, $H$=mean pairwise normalized Hamming distance, $\rho$=latent-std ratio, $S_{\mathrm{noise}}$=noise sensitivity.}
\end{table}

\begin{table}[t]
	\caption{Diversity indicators (round-1 attack), Drebin ($M{=}215$).}
	\label{tab:div215}
	\centering
	\begin{tabular}{|c|c|c|c|c|}
		\hline
		\textbf{Detector} & \textbf{$U$} & \textbf{$H$} & \textbf{$\rho$} & \textbf{$S_{\mathrm{noise}}$} \\
		\hline
		DT & 1.00 & 0.20 & 1.41 & 0.36\\\hline
		KNN & 0.95 & 0.16 & 1.12 & 0.06\\\hline
		GB & 0.61 & 0.07 & 3.44 & 0.50\\\hline
		AB & 0.72 & 0.07 & 3.24 & 0.65\\\hline
		MLP & 0.75 & 0.09 & 3.21 & 0.21\\\hline
		RF & 0.78 & 0.10 & 5.05 & 0.79\\\hline
		SVM & 0.75 & 0.10 & 4.99 & 0.98\\\hline
		LR & 0.84 & 0.12 & 1.63 & 0.10\\\hline
	\end{tabular}
	\vspace{4pt}

	{\footnotesize Note: $U$=unique ratio, $H$=mean pairwise normalized Hamming distance, $\rho$=latent-std ratio, $S_{\mathrm{noise}}$=noise sensitivity. Across detectors $U$ ranges $0.61$--$1.00$ and $\rho\gtrsim1$.}
\end{table}

On Drebin every model adds more API features than on CICMaldroid 2020 because more features must be added to cross the decision boundary and the Drebin feature space is larger, but Mal-AAEGAN remains the most economical generator there and the most noise-responsive on both datasets; its unique ratio ($0.799$) sits above Mal-D2GAN's and below Mal-LSGAN's and MalGAN's.

\subsection{Discussion: perturbation cost and noise sensitivity}
The comparison in Table~\ref{tab:modelcmp} highlights a difference that a raw evasion reading hides. On both datasets, the four generators reach a similarly high evasion (mostly $94$--$100\%$) while adding a similar number of API features (about $21$--$51$ features on CICMaldroid 2020, $43$--$92$ on Drebin); Mal-AAEGAN reaches the highest evasion on CICMaldroid 2020 and is among the most economical models, adding the fewest features on Drebin and fewer than MalGAN and Mal-LSGAN on CICMaldroid 2020. Since every model keeps the number of added features modest and the OR pipeline preserves functionality~\cite{b25}, the models separate most clearly not on evasion or cost but on how the generator uses its noise input.

The two diversity indicators refine this picture. The noise sensitivity of all three baselines sits one to three orders of magnitude below that of Mal-AAEGAN ($S_{\mathrm{noise}}\approx10^{-3}$--$10^{-2}$ versus $0.075$ on CICMaldroid 2020 and $0.411$ on Drebin), so the baseline generators barely respond to their noise input and produce essentially one adversarial variant per malware, while Mal-AAEGAN's output varies with the noise. The unique ratio is more moderate for Mal-AAEGAN ($0.320$ on CICMaldroid 2020 and $0.799$ on Drebin): the additive latent perturbation combined with the decode-binarize-OR pipeline maps a share of the latent codes back to similar binary vectors, an effect that is stronger on the lower-dimensional CICMaldroid 2020 feature space. We report both $U$ and $S_{\mathrm{noise}}$ as descriptive measures of the generated set rather than as pass/fail criteria.

For the defensive use case that motivates this line of work -- red-teaming and hardening detectors -- the noise-responsive generator is the property of practical interest: at a comparable perturbation cost it yields an attack set that is not reducible to a single perturbation per malware, giving broader coverage when the generated samples are used to retrain a detector.

\section{Conclusion and Future Work}\label{sec:conc}
We presented Mal-AAEGAN, an adversarial-malware generator that combines an adversarial autoencoder with a double-critic WGAN-GP. Moving the generator from the discrete $M$-dimensional binary space into a continuous, prior-regularized latent space, and replacing the least-square detectors with two Wasserstein critics, is intended to stabilize training and reduce the perturbation cost of the generated adversarial samples while preserving the malware functionality through the decode-binarize-OR pipeline. The model is evaluated on two malware datasets of different input dimensionality ($M\in\{215,139\}$) against eight detectors and compared with MalGAN, Mal-LSGAN and Mal-D2GAN under an identical protocol. Mal-AAEGAN evades on par with or better than the binary-space baselines -- reaching the highest evasion of all four models on CICMaldroid 2020 and a $98.9\%$ mean evasion rate on Drebin -- at a comparable perturbation cost (the fewest added API features of the four models on Drebin), and its generator is markedly more responsive to the noise input; the unique ratio of its samples is moderate on both datasets.

Evaluating the robustness of the generated samples against adversarially retrained detectors is beyond the scope of this paper and is left to future work.

Future work will study joint fine-tuning of the AAE with the WGAN-GP stage, alternative priors for the latent space, and validation across more diverse datasets.

\begin{credits}

\end{credits}
%
%
%
%

\end{document}